\documentclass[aps,prl,twocolumn,showpacs]{revtex4}
\usepackage{amssymb}
\usepackage{amsmath}
\usepackage{epsfig}

\begin{document}

\title{Gap soliton dynamics in an optical lattice as a parametrically
driven pendulum}
\author { R. Khomeriki${}^{1,2}$,  J. Leon${}^1$}
\affiliation {
(${\ }^1$) Laboratoire de Physique Th\'eorique et Astroparticules \\
  CNRS-IN2P3-UMR5207, Universit\'e Montpellier 2, 34095 Montpellier (France)\\
(${\ }^2$)  Physics Department, Tbilisi State University, 0128
  Tbilisi (Georgia)}

\begin{abstract}  A long wavelength optical lattice is generated in a two-level
medium by low-frequency contrapropagating beams. Then a short wave length gap
soliton generated by evanescent boundary instability (supratransmission)
undergoes a dynamics shown to obey the Newton equation of the parametrically
driven pendulum, hence presenting extremely rich, possibly chaotic, dynamical
behavior.  The theory is sustained by numerical simulations and provides an
efficient tool to study soliton trajectories.
\end{abstract}  \maketitle

\paragraph{Introduction.}

Nonlinear physics has revealed that quite different complex systems may
actually share the same model equations with common simple and universal
physical properties \cite{scott}. One celebrated example is the
Fermi-Pasta-Ulam chain of anharmonic oscillators \cite{fpu}, which may serve as
a laboratory to check soliton theory, statistical physics and even dynamical
processes in DNA molecules. Another famous example is the Josephson junction,
mathematically analogue to a pendulum, where the biased voltage and the internal
resistance play the role of forcing and damping \cite{dyn-bif1}.

\begin{figure}[t] \centerline {\epsfig{file=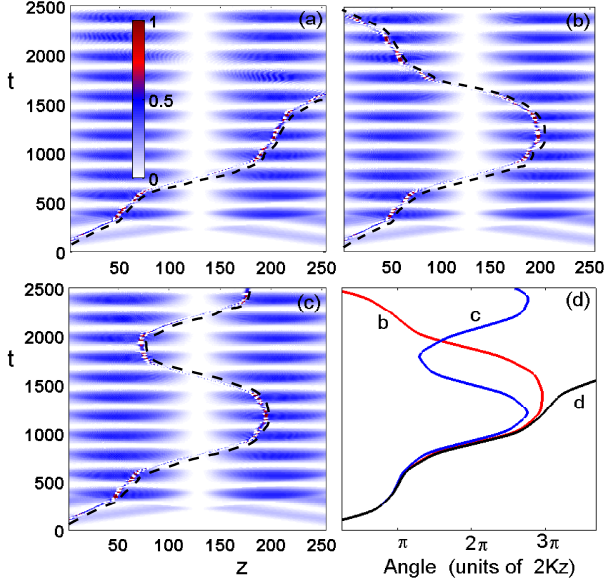,width=\linewidth}}
\caption{(Color online) \it Graphs (a), (b) and (c) display results of numerical
simulation of (\ref{MB}) under boundary conditions (\ref{parameters}) for the
three incident wave amplitudes (\ref{abc}) generating optical lattice with
slightly different depths. Graph (d) shows the pendulum angle evolution in model
(\ref{parapend}) corresponding to those three regimes. For sake of presentation,
the gap soliton trajectories have been underlined with dashed lines.}
\label{fig:surf1}
\end{figure}

We consider here another well established Maxwell-Bloch (MB)
model \cite{Fain} which has an universal character related to
various nonlinear processes \cite{jerome,ref} in nonlinear optics,
and discover a new example of a reduction of a complex many body
dynamics of MB system to the driven-damped pendulum motion. This
is done in the context of gap soliton motion in a two level medium
subjected to some low frequency stationary boundary driving. We
show that the soliton trajectory in this effective optical lattice
(periodic in space and time) obeys the equation of a parametric
pendulum which results in a rich dynamical behavior, from periodic
to chaotic, depending on the fundamental parameters of the problem
\cite{pend1,pend2}. Fig.\ref{fig:surf1} displays three instances
of the gap soliton propagation in a two level medium prepared as
an optical lattice. The trajectories compare well to the time
dependence of the angle of a parametric pendulum.

In a two-level system (TLS) of transition frequency $\Omega_0$,
the governing equation is the MB model \cite{Fain}, considered
here in the isotropic case for a linearly polarized
electromagnetic field propagating in direction $z$. The time is
scaled to the inverse transition frequency $\Omega_0^{-1}$, the
space $z$ to the length $\Omega_0 c/\eta$ with optical index
$\eta=\sqrt{\epsilon\mu_{_0}}$, the energy to the average
$W_0=N_0\hbar\Omega_0/2$, the electric field to
$\sqrt{W_0/\epsilon}$ and the polarization to $\sqrt{\epsilon
W_0}$. Here $N_0$ is the density of active dipole. The resulting
dimensionless MB system then reads
\begin{equation}
(E+P)_{tt}=E_{zz},\quad
P_{tt}+P=-\alpha NE,\quad
N_t=EP_t.\label{MB}
\end{equation}
The electric field is $\textbf{E}=(E(z,t),0,0)$ and the polarization source
$\textbf{P}=(P(z,t),0,0)$. The quantity $N(z,t)$ is the \textit{normalized
inversion of population density} that is assumed to be $-1$  in the fundamental
state (no applied field). The coupling  strength is the dimensionless
fundamental constant $\alpha$ completely characterized by the gap opening
between the transition frequency (value $1$ in reduced units) and the plasma
frequency $\omega_0$. This gap results from the linear ($N=-1$) dispersion
relation of the MB system
\begin{equation}\label{disp}
\omega^2(\omega^2-\omega_0^2)=k^2(\omega^2-1),\quad \omega_0^2=1+\alpha,
\end{equation}
obtained for a carrier $\exp[i(\omega t-kz)]$.

Our purpose is to create an optical lattice with two contra-propagating beams of
frequency $\Omega\ll 1$, which then will interact with a wavepacket of central
frequency $\omega\le\omega_0$ (inside the gap, close to the upper gap edge).
The contra-propagating beams are expected to create a stationary wave in the
variables $E$, $P$ and $N$ that will then interact with the gap soliton
through mediation of the two fundamental coupling terms $NE$ and $EP_t$ in MB
equations. To achieve this study we shall derive from (\ref{MB}) a
limit model by the reductive perturbative expansion method where
essential phase effects are carefully taken into account \cite{oikawa}.
We shall obtain the following dynamical equation for the soliton motion, in some
newly normalized time $t$ and position $q(t)$:
\begin{equation}\label{parapend}
\ddot{q}+\gamma\dot{q}= -A\sin^2(t)\,\sin(q), \quad
A={\cal E}^2\frac{\alpha^2(2-\alpha)}{2(1+\alpha)},
\end{equation}
which is a parametrically driven pendulum. Here ${\cal E}$ is the amplitude of
the stationary standing wave and $\gamma$ is a small phenomenological damping
parameter accounting for soliton energy losses trough the optical lattice. The
comparison of this effective dynamics with the numerical simulations on the full
MB model is presented in fig.\ref{fig:surf1}.

\paragraph{Theory.}

We consider an electric field that carries two fundamental
frequencies: one is close to the band gap edge $\omega_0$, the other is close
to zero. Then one deals with a two-wave nonresonant interaction process whose
weakly nonlinear limit is sought by assuming the formal series expansion
\cite{oikawa}
\begin{eqnarray}
F=\sum\limits_{p=1}^\infty\sum\limits_{\ell=-p}^p
\varepsilon^p F_{\ell}^{(p)}(\xi_1,\xi_2,\tau)
e^{i\ell[\omega_0 t+\theta(\xi_2,\tau)]},\nonumber \\
\xi_1=\varepsilon\left[z-\phi(\xi_2,\tau)\right], \quad \xi_2=\varepsilon^2z,
\quad \tau=\varepsilon^2 t, \label{multi}
\end{eqnarray}
where $F$ stands for any of the three fields $E(z,t)$, $P(z,t)$ or
$n(z,t)=1+N(z,t)$. Note that now $n(z,t)$ denotes the normalized
population of the \textit{excited state}. As all components are
real valued, we have $\bar F_{\ell}^{(n)}=F_{-\ell}^{(n)}$
(overbar stands for complex conjugate). The slow space variables
$\xi_1$ and $\xi_2$ are associated respectively with the
characteristic wavelengths of the gap soliton and of the standing
wave grating. The above representation actually means to replace
the differential operators as follows
\begin{align}
& \frac{d}{dz}\,\to\, \varepsilon\frac{\partial}{\partial\xi_1}+
\varepsilon^2\frac{\partial}{\partial\xi_2}
-\varepsilon^3\frac{\partial\phi}{\partial\xi_2}\frac{\partial}{\partial\xi_1},
\nonumber\\
& \frac{d}{dt}\,\to\, i\ell\omega_0+\varepsilon^2\frac{\partial}{\partial\tau}
-\varepsilon^3\frac{\partial\phi}{\partial\tau}\frac{\partial}{\partial\xi_1},
\end{align}
when applied to a given harmonic $\ell$ of the asymptotic series inserted in
(\ref{MB}). This provides the following first order structure of the field
\begin{align}
E= &\varepsilon \big[ E_{1}^{(1)}(\xi_1,\tau)
e^{i[\omega_0 t+\theta(\xi_2,\tau)]} +c.c.\big] \nonumber \\
&+\varepsilon E_{0}^{(1)}(\xi_2,\tau)+{\cal O}(\varepsilon^2), \nonumber \\
P= & -\varepsilon\big[ E_{1}^{(1)}(\xi_1,\tau)
e^{i[\omega_0 t+\theta(\xi_2,\tau)]} +c.c.\big] \nonumber \\
& + \varepsilon\alpha E_{0}^{(1)}(\xi_2,\tau)+{\cal O}(\varepsilon^2), \nonumber
\\
n=&{\cal O}(\varepsilon^2),\label{linear}
\end{align}
which accounts for the linear limit and deserves some comments. The set of
unknowns is:  the slowly varying envelope $E_{1}^{(1)}(\xi_1,\tau)$ of the short
wavelength wavepacket (e.g. gap soliton), the \textit{very slowly} varying
profile $E_{0}^{(1)}(\xi_2,\tau)$  of the long wavelength applied grating, the
phase $\theta(\xi_2,\tau)$ which accounts for variations due to wave-coupling
and finally $\phi(\xi_2,\tau)$ which will be related to the position of the
wavepacket by means of (\ref{multi}). Then we note that the expression of the
polarization in terms of the electric field actually does follow the usual
linear laws $P=-E$ for the plasma wave (here given by $E_{1}^{(1)}e^{i\omega_0
t}$, the mode at $\omega=\omega_0$) and $P=\alpha E$ for the electrostatic
component (here given by $E_{0}^{(1)}$, the mode at $\omega=0$). The dependence
of these two fields in the slow variables, together with phase variations, will
then carry the nonlinear electrodynamics.

Next we seek the leading order ($\varepsilon^2$) harmonics of $n(z,t)$.
According to the last equation of (\ref{MB}), using (\ref{linear}) and
collecting harmonics $\ell= 1$ and $\ell= 2$, we get after time-integration
\begin{equation}
n^{(2)}_{1}=-E_1^{(1)}E_{0}^{(1)}, \quad n^{(2)}_{2}=
-\frac{1}{2}\big(E_{1}^{(1)}\big)^2 .\label{n2int}
\end{equation}
The last term to compute is $n^{(2)}_{0}$ which cannot be calculated at order
$\varepsilon^2$ since there, the third equation of (\ref{MB}) is trivial. Thus
we move one step further (actually to $\varepsilon^4$ to catch the
$\tau$-derivative), which provides
\begin{align}\label{n00}
\frac{\partial n^{(2)}_{0}}{\partial\tau}&=- E_{1}^{(1)}
\frac{\partial\bar E_{1}^{(1)}}{\partial \tau}-
\bar E_{1}^{(1)}\frac{\partial E_{1}^{(1)}}{\partial \tau}
+\alpha E_{0}^{(1)} \frac{\partial E_{0}^{(1)}}{\partial \tau}\nonumber\\
&+i\omega_0\left[(E ^{(3)}_{1}+P^{(3)}_{1})\bar E^{(1)}_{1}-c.c.\right].
\end{align}
Then writing the second equation of (\ref{MB}) at order $\varepsilon^2$ and
using  (\ref{n2int}), we obtain for $\ell=1$
\begin{align}\label{third}
&\alpha(P^{(3)}_{1}+ E^{(3)}_{1})+ 2i\omega_0\frac{\partial
E^{(1)}_{1}}{\partial\tau}-2\omega_0E^{(1)}_{1}\frac{\partial
\theta}{\partial\tau} \\ &-\frac{4+\alpha}{2}
\big|E_{1}^{(1)}\big|^2E_{1}^{(1)} +\alpha\big(E_{0}^{(1)}\big)^2
E_{1}^{(1)}-\alpha n^{(2)}_{0}E_{1}^{(1)}=0, \nonumber
\end{align}
which is replaced in (\ref{n00}). The result can be integrated
to furnish the sought expression
\begin{equation}\label{n00int}
n^{(2)}_{0}=\frac{2+\alpha}{\alpha} \left|E_{1}^{(1)}\right|^2
+\frac{\alpha}{2} \big(E_{0}^{(1)}\big)^2.
\end{equation}
Next we consider first equation of (\ref{MB}) at order $\varepsilon^3$
and collect again the $\ell=1$ harmonics to get
\begin{equation}\label{three}
\omega_0^2\big(E^{(3)}_{1}+ P^{(3)}_{1}\big)+\frac{\partial^2
E^{(1)}_{1}}{\partial\xi_1^2}=0.
\end{equation}
After some simple algebra we can eliminate $E^{(3)}_{1}$ and $P^{(3)}_{1}$ from
equations (\ref{third}) and (\ref{three}). The resulting equation appears
with terms that depend solely on a single variable,  $E^{(1)}_{1}(\xi_1,\tau)$
on the one side, $\theta(\xi_2,\tau)$ and $E_{0}^{(1)}(\xi_2,\tau)$ on the other
side. These terms thus decouple to eventually give
\begin{align}
&2i\omega_0\frac{\partial E^{(1)}_{1}}{\partial\tau}=
\frac{\alpha}{\omega_0^2}\frac{\partial^2
E^{(1)}_{1}}{\partial\xi_1^2}+\frac{4+\alpha}{2}\big|E_{1}^{(1)}\big|^2
E_{1}^{(1)},\label{NLS}\\
&\frac{\partial\theta}{\partial\tau}=-
\frac{\alpha(\alpha-2)}{4\omega_0}\big(E_{0}^{(1)}\big)^2.\label{sch3}
\end{align}
It appears therefore that $E^{(1)}_{1}(\xi_1,\tau)$ obeys a nonlinear
Schr\"odinger equation where the effect of the applied grating $E_{0}^{(1)}$
lies in the definition of the variable $\xi_1$ in (\ref{multi}). Thus
the drift $\phi(\xi_2,\tau)$ remains to be evaluated, which is done
with the last equation of (\ref{MB}) at order $\varepsilon^4$ and $\ell=1$
\begin{equation}\label{fin}
\frac{\partial\phi}{\partial\tau}=-\frac{\alpha}{\omega_0^3}
\frac{\partial\theta}{\partial\xi_2}.
\end{equation}
Here we have assumed  $E^{(2)}_1=0$ which is allowed by the structure of
(\ref{MB}) where nonlinearity comes into play at order $\varepsilon^3$. Indeed
one actually obtains a linear homogeneous equation for $E^{(2)}_1$ whose unique
solution is $E^{(2)}_1=0$ as soon as the initial-boundary value problem
concerns only $E^{(1)}_{1}$.

To close the system (\ref{NLS})-(\ref{fin}) we derive the equations for the
low frequency oscillations obtained from the first equation of (\ref{MB}) at
$\ell=0$, namely
\begin{equation}\label{wave}
\omega_0^2\,\frac{\partial^2 E^{(1)}_{0}}{\partial\tau^2}-
\frac{\partial^2 E^{(1)}_{0}}{\partial{\xi_2}^2}=0.
\end{equation}
The system (\ref{NLS})-(\ref{wave}) is the basic set of equations that describes
within the Maxwell-Bloch model, the interaction of a gap soliton with a long
wavelength standing wave. Note that all coefficients are completely determined
from the unique parameter $\alpha$, the fundamental coupling constant of
eq.(\ref{MB}), as by definition $\omega_0^2=1+\alpha$. It is useful to eliminate
the phase $\theta$ between (\ref{sch3}) and (\ref{fin}) and obtain
\begin{equation}\label{phi}
\frac{\partial^2\phi}{\partial\tau^2}=
\frac{\alpha^2(\alpha-2)}{4(1+\alpha)^2}
\frac{\partial}{\partial\xi_2}\big(E_{0}^{(1)}\big)^2,
\end{equation}
which constitutes with (\ref{wave}) a closed system, \textit{independently}
of $E^{(1)}_{1}$.

\paragraph{Application.}

We apply now the above machinery to the case when the fundamental
field component $E^{(1)}_{1}$ is a gap soliton, exact solution to
(\ref{NLS}), of given velocity $v_0$ in the frame $(\xi_1,\tau)$.
The nature of equation (\ref{NLS}) allows us to start with such an
explicit solution and then to study its dynamics by looking at the
frame drift $\phi(\xi_2,\tau)$ which thus defines the variations
of the soliton about its free motion. To that end it is more
convenient to rewrite system (\ref{wave})(\ref{phi}) in the
physical (dimensionless) variables, namely
\begin{align}
 &\omega_0^2\,\frac{\partial^2 E_0}{\partial t^2}-
\frac{\partial^2 E_0}{\partial z^2}=0,\nonumber\\
&\frac{\partial^2\phi}{\partial t^2}=
\frac{\alpha^2(\alpha-2)}{4(1+\alpha)^2}
\frac{\partial}{\partial z}\big(E_0\big)^2,
\end{align}
where $E_0(z,t)=\varepsilon E_0^{(1)}(\xi_2,\tau)$. Now the soliton is localized
in $\xi_1$ and its position, say $z_s(t)$, is defined by $\xi_1={const}$,
namely by $z_s-\phi(z_s,t)={const}$. Computing the velocity $dz_s/dt$ and the
acceleration $d^2z_s/dt^2$ (total derivatives) we remember that $\phi$ is slowly
varying in $z$ and $t$ and thus keep only the dominant orders to eventually
obtain
\begin{equation}\label{acce-sol}
\frac{d^{\,2}z_s}{dt^2}\simeq
\left(\frac{\partial^2\phi}{\partial t^2}\right)_{z=z_s}.
\end{equation}

We may now select a particular solution to the linear wave
equation (\ref{wave}) that would result from application to the
medium of two contrapropagating monochromatic beams. The effective
boundary conditions that represent such a situation will be
described later, they result in the generation of a standing wave
at frequency $\Omega$, namely (assuming a system length
corresponding to a mode at that frequency)
\begin{equation}\label{grating}
E_{0}={\cal E}\sin(\Omega t)\sin(Kz), \quad
K^2=(1+\alpha)\Omega ^2.
\end{equation}
Note that the above dispersion law is indeed the behavior at small $\omega$
of the general dispersion law (\ref{disp}). The resulting equation for the
soliton acceleration (\ref{acce-sol}) then reads
\begin{equation}\label{acc}
\frac{d^{\,2}z_s}{dt^2}=
K{\cal E}\frac{\alpha^2(\alpha-2)}{4(1+\alpha)^2}
\,\sin^2(\Omega t)\,\sin(2K z_s).
\end{equation}
This is the parametric driven pendulum equation that can be
written as (\ref{parapend}) by rescaling time ($\Omega t\to t$)
and position ($2Kz_s=q$), and by adding a phenomenological damping
to account for soliton radiation. Such an equation is solved with
the data of initial soliton position  $z_0$ and velocity  $v_0$.
Our purpose now is to compare the solution of Maxwell-Bloch
(\ref{MB}) with convenient boundary data to the prediction of
soliton trajectory given by (\ref{parapend}).

\paragraph{Numerical simulations.}

To  proceed with numerical simulations of the MB equations (\ref{MB}) we derive
the boundary-value data following  \cite{chen}, that represent incident
waves entering the system at $z=0$ and $z=L$ and which are expected to produce
the standing low frequency wave (\ref{grating}). The vacuum outside $[0,L]$ is
assumed to obey (\ref{MB}) with $\alpha=0$ whose solution reads
\begin{equation}\label{z0}
z\le0\,:\quad E_{vac}=I\cos[\Omega(t-z)]+R\cos[\Omega(t+z)].
\end{equation}
The amplitude $I$ of the incident wave from the left is the control parameter,
the amplitude $R$ of the reflected wave has thus to be eliminated. The
continuity conditions at $z=0$ with the electric field $E(z,t)$ inside the
medium can be written
\begin{align*}
(I+R)\cos(\Omega t)= \big(E\big)_{z=0}, \\
\Omega(I-R)\sin(\Omega t)=\big(\partial_zE\big)_{z=0},
\end{align*}
which can be combined to eliminate the unknown reflected amplitude
$R$ to give
\begin{align*}
\big(\partial_zE-\partial_tE\big)_{z=0}=2\Omega I\sin(\Omega t),\\
\big(\partial_zE+\partial_tE\big)_{z=L}=-2\Omega Q\sin(\Omega t).
\end{align*}
The above second relation is obtained at t $z=L$ and $Q$ is the chosen amplitude
of the incident wave from the right. Such unusual boundary data are actually
numerically implemented in a finite difference scheme as (define $L^-=L-dz$)
\begin{align*}
&E(0,t)=E(dz,t)-dz\,[\partial_tE(dz,t)+2\Omega I\sin(\Omega t)],\\
&E(L,t)=E(L^-,t)-dz\,[\partial_tE(L^-,t)+2\Omega Q \sin(\Omega t)],
\end{align*}
which are inserted in the differential equation (\ref{MB}). We shall
use equal amplitude out of phase driving, namely $I=-Q$. This
generates the standing wave solution (\ref{grating}) whose amplitude
${\cal E}$ is then defined from $I$ by
\begin{equation}\label{standing}
{\cal E}=2I\left[ (1+\alpha)\cos^2(KL/2)+\sin^2(KL/2)\right]^{-1/2}.
\end{equation}

Then the gap soliton is generated by driving the boundary $z=0$ with
an incident pulse at frequency $\omega$ \cite{sls} and finally the
full set of boundary conditions that produce the
plots (a), (b) and (c) in fig.\ref{fig:surf1} read
\begin{align}\label{parameters}
\big(\partial_zE-\partial_tE\big)_{z=0}&=2I\Omega \sin(\Omega t)
+\frac{1.393\cos(\omega t)}{\cosh[(t-100)/12]} ,
\nonumber \\
\big(\partial_zE+\partial_tE\big)_{z=L}&=2I\Omega \sin(\Omega t),
\end{align}
for a system length $L=254$ and initial state at rest: $E(z,0)=0$, $P(z,0)=0$,
$N(z,0)=-1$. We choose $\alpha=1$, $\Omega =\pi/200$ and $\omega=1.5$. As
defined by (\ref{grating}) the generated standing wave has wavenumber
$K=\Omega\sqrt{1+\alpha}$ and its amplitude is defined by (\ref{standing}). The
only parameter we vary is the amplitude $I$ of the contrapropagating low
frequency beams. In particular the graphs (a), (b) and (c) of
fig.\ref{fig:surf1} correspond to the following values
\begin{equation}\label{abc}
2I_a=1.832, \qquad 2I_b=1.843, \qquad 2I_c=1.848.
\end{equation}
Why such a small change of the control parameter causes completely different
trajectories of the soliton is understood using the pendulum description
(\ref{parapend}) where in view of (\ref{standing}) the amplitude $A$ is related
to the parameters of MB (with $\alpha=1$) by
\begin{equation}\label{para}
A=I^2\left[2\cos^2(KL/2)+\sin^2(KL/2)\right]^{-1}.
\end{equation}
We now solve (\ref{parapend}) with initial conditions $z_s(0)=0.8$ and
$v_0=0.94$ that have been measured as the common gap soliton initial position
and velocity in the three successive simulations (a), (b) and (c). The
phenomenological damping constant is $\gamma=0.01$. Then the pendulum angle
evolution for three different values of $A$ resulting from (\ref{para}) and
(\ref{abc}) are displayed in the graph (d) of fig.\ref{fig:surf1}.

\paragraph{Conclusion and comments.}

It is remarkable that such different models as the Maxwell-Bloch
system (\ref{MB}) and the driven parametric pendulum
(\ref{parapend}) concur to describe the dynamics of a gap soliton
interacting with a low-frequency standing wave. This demonstrates
in particular that the gap soliton trajectory may well be chaotic,
at least initially (it may stabilize due to energy losses by
radiationor the internal damping of MB model). This is a direct
result of the time dependence of the grating induced by the
stationary field. In particular the gap soliton dynamics in a
permanent grating (as e.g. a Bragg medium) is regular.

It is worth noting that the effect of damping in initial MB model
(\ref{MB}), originating from finite dephasing times,  causes
eventual transition from chaotic  to self trapped regime, as
expected from a driven damped pendulum, and which we have checked
on numerical simulations. This occurs for time scales of the order
of the dephasing times, for which the soliton gets trapped into
one of the nodes of the standing wave grating.

One practical advantage of the parametric pendulum description is
the prediction of the switching from regular to chaotic (erratic)
motion of the gap soliton in terms of the parameters of the
driving field. However we expect that the various scenario of
chaotic versus periodic motion could be more complex than the
driven pendulum case and a more comprehensive analysis would first
require much longer computation times, which is a subject of
further studies. Beyond such a technical result, our analysis has
revealed a new type of soliton motion in a medium with dynamically
generated optical grating.

We expect to extend this analysis of a one dimensional time-dependent
situation to the study of spatial soliton trajectories in two dimensional
smooth optical lattices.

\paragraph{Acknowledgements.} Work done under contract CNRS GDR-PhoNoMi2
(\textit{Photonique Nonlin\'eaire et Milieux Microstructur\'es}).
R.K. aknowledges stay as invited professor at the
\textit{Laboratoire de Physique Th\'eorique et Astroparticules}
and financial support of the Georgian National Science Foundation
(Grant No GNSF/STO7/4-197).


\begin{thebibliography}{aa99}

\bibitem{scott} A.C. Scott, {\it Nonlinear Science}, 2-nd edition.
(Oxford University Press, New York, 2003).
\bibitem{fpu} E. Fermi, J. Pasta, S. Ulam, M. Tsingou, in
{\it The Many-Body Problems}, edited by
D.C. Mattis (World Scientific,  Singapore, 1993 reprinted); G.
Gallavotti (Ed.) {\it The Fermi-Pasta-Ulam Problem: A status
report}, Springer (2008).
\bibitem{dyn-bif1} I. Siddiqi, R. Vijay, F. Pierre, C.M. Wilson, L.
Frunzio, M. Metcalfe, C. Rigetti, R.J. Schoelkopf, M.H. Devoret,
D. Vion, D. Esteve, Phys Rev Lett 94 (2005) 027005
\bibitem{Fain} V.M. Fain and Ya.I. Khanin (1969) \textit{Quantum Electronics},
Pergamon (Oxford, UK)
\bibitem{jerome} R. Khomeriki, J. Leon, Phys Rev Lett 99 (2007) 183601.
\bibitem{ref} M. Clerc, P. Coullet, E. Tirapegui, Opt. Communications 167 (1999) 159
\bibitem{pend1} J. Starrett, R. Tagg, Phys Rev Lett 74 (1995) 1974
\bibitem{pend2} A.D. Churukian, D.R. Snider, Phys Rev E 53 (1996) 74
\bibitem{oikawa} M. Oikawa, N. Yajima, J. Phys. Soc. Japan 37 (1974) 486
\bibitem{sls} J. Leon, Phys Rev A 75 (2007) 063811
\bibitem{chen} W. Chen, D.L. Mills, Phys Rev B, 35 (1987), 524
\end{thebibliography}
\end{document}